\documentclass[preprint,aps]{revtex4}
\usepackage{graphicx,amsfonts,amsmath}% Include figure files
\usepackage{dcolumn}% Align table columns on decimal point
\usepackage{bm}% bold math

\newcommand{\pa}{\partial}                           %new

\newcommand{\be}{\begin{equation}}
\newcommand{\ee}{\end{equation}}
\newcommand{\bea}{\begin{eqnarray}}
\newcommand{\eea}{\end{eqnarray}}

\def\pa{\partial}

\begin{document}

\title{On the noncommutative fields method in the three-dimensional Yang-Mills theory}

\author{J. R. Nascimento, A. Yu. Petrov, E. O. Silva}
\affiliation{Departamento de F\'{\i}sica, Universidade Federal da Para\'{\i}ba\\
Caixa Postal 5008, 58051-970, Jo\~ao Pessoa, Para\'{\i}ba, Brazil}
\email{jroberto,petrov,edilberto@fisica.ufpb.br}

\begin{abstract}
We apply the noncommutative fields method to the three-dimensional non-Abelian gauge theory. We find that, first, implementing the noncommutativity between the canonical momenta implies in generation of the non-Abelian Chern-Simons term, second, if one introduces the noncommutativity between the field operators, the higher derivative terms would arise.
\end{abstract}

\maketitle
\newpage

The noncommutativity is treated now as a fundamental quantum property of the space-time geometry. Beside of the known scheme of introducing the noncommutativity via the Moyal product \cite{SW}, an alternative one was recently developed, that is, so-called noncommutative fields method, in which, instead of the spacetime coordinates, fields themselves are noncommutative, thus, the canonical commutation relations turn out to be deformed \cite{Gamb0}. This method turned out to be a new method of generating the Lorentz-breaking correction after it was shown that the known Lorentz-breaking term initially introduced by Jackiw and Kostelecky \cite{JK} naturally emerges within this formalism \cite{Gamb1}. Further, the non-Abelian analog of this term was generated via the noncommutative fields method \cite{Gamb2}, and in our paper \cite{ourgra}, this method was applied to generate the Lorentz symmetry breaking in the linearized gravity. 

At the same time, the situation in three-dimensional space-time is different. Indeed, we have shown in \cite{NPR} that application of the noncommutative field method to three-dimensional electrodynamics, instead of the Lorentz-breaking terms generates a gauge invariant mass term, that is, the Chern-Simons term, with the mass turns out to be proportional to the noncommutativity parameter \cite{NPR}.  We would like to notice that unlike of common perturbative approach (see f.e. \cite{Redlich}), the essence of the noncommutative fields method consists in possibility to generate new terms without coupling to extra matter fields. The very natural development of this study would consist in generalization of the noncommutative fields method for the non-Abelian case, where it is natural to expect that not only quadratic term but also the interaction term for the gauge field will arise. 
Different aspects of the Chern-Simons term, both in Abelian and non-Abelian cases, such as non-trivial topological nature of this term \cite{DJT} and quantization of the Chern-Simons coefficient \cite{quCS} were studied.
In other worlds, it is natural to expect that in this case, the three-dimensional non-Abelian Chern-Simons term
\bea
L_{CS}=\frac{1}{2}m\epsilon^{\mu\nu\lambda}{\rm tr}(A_{\mu}\pa_{\nu}A_{\lambda}+\frac{2}{3}gA_{\mu}A_{\nu}A_{\lambda})
\eea
will be generated. 
From the other side, we are planning to generalize the noncommutative field method by introducing of a more general deformation of the canonical algebra which in principle could imply in arising of the Lorentz-breaking terms. These problems are considered in the paper.

Let us start our study of the three-dimensional Yang-Mills theory, whose action is
\bea
S=-\frac{1}{4}\int d^3x {\rm tr}F_{mn}F^{mn},
\eea
with the $F_{mn}=F_{mn}^aT^a$ is a stress tensor constructed on the base of the Lie-algebra valued gauge field $A_m(x)=A_m^a(x)T^a$ (with ${\rm tr}(T^aT^b)=\delta^{ab}$, and $[T^a,T^b]=f^{abc}T^c$):
\bea
F_{mn}^a=\pa_m A_n^a-\pa_n A_m^a+gf^{abc}A^b_mA^c_n,
\eea
so, the Lagrangian, after splitting of the indices into time (zero) and space ones (denoted by $i,j,k$) looks like
\bea
L=-\frac{1}{4}\int d^3x F_{mn}^aF^{mn\,a}=-\frac{1}{4}F^a_{ij}F^a_{ij}+\frac{1}{2}(\dot{A}^a_i-\pa_i A^a_0+gf^{abc}A^b_iA^c_0)^2.
\eea
Let the signature be $diag(-++)$.
First, we carry out the canonical quantization of the theory. The canonical momentum of the theory is
\bea
\label{cm0}
p^a_m=\frac{\pa L}{\pa \dot{A}^{a\,m}}=F^a_{0m}.
\eea
It is clear that $p^a_0=0$, so, we find the primary constraint $\Phi^{(1)a}=p^a_0$. The velocities can be expressed as
\bea
\label{vel}
\dot{A}^a_i=p_i^a-gf^{abc}A^b_0A^c_i+\pa_i A_0^a.
\eea
Thus, the Hamiltonian is
\bea
H=p^a_i\dot{A}^a_i-L=\frac{1}{2}p^a_ip^a_i+\frac{1}{4}F^a_{ij}F^a_{ij}+p_i^a(-gf^{abc}A^b_0A^c_i+\pa_i A_0^a).
\eea
The secondary constraint looks like
\bea
\Phi^{(2)b}\equiv \Delta^a=\{p^a_0,H\}=-\frac{\pa H}{\pa A_0^a}=-(\pa_i p_i^a+gf^{abc}A^b_i p_i^c)\equiv
-{\cal D}^{ab}_ip^b_i.
\eea
This constraint evidently generates the gauge transformations:
\bea
\label{gatra}
\delta A_i^a&=&\{A_i^a, \int d^2\vec{x}\xi^b(\vec{x})\Delta^b(\vec{x})\}=\pa_i\xi^a(\vec{x})+gf^{abc}A^b_i(\vec{x})\xi^c(\vec{x})\,(\equiv{\cal D}^{ac}_i\xi^c(\vec{x}));\nonumber\\
\delta p_i^a&=&\{p_i^a, \int d^2\vec{x}\xi^b(\vec{x})\Delta^b(\vec{x})\}=-gf^{abc}\xi^b(\vec{x})p^c_i(\vec{x}),
\eea
which evidently reproduces the known gauge transformation for the connection and stress tensor. Here the ${\cal D}^{ac}$ is a gauge covariant derivative.

It is easy to check that the primary and secondary constraints mutually commute, $\{\Phi^{(1)a},\Phi^{(2)b}\}=0$. Further, one can find that $\{\Phi^{(2)b},H\}=0$, thus, no new constraints arise (see also \cite{Park,Wo} for discussion of the canonical structure of the theories with the Chern-Simons term).

The canonical quantization of the theory can be carried out in a standard way, that is, we define the canonical variables $A^a_i$ and $p^a_i$ to be operators with the commutation relation $[A^a_i(\vec{x}),p^b_j(\vec{y})]=i\delta_{ij}\delta^{ab}\delta(\vec{x}-\vec{y})$, with all other commutators of the canonical variables be zero.

Now, let us implement the noncommutative fields method. To do it, we deform the canonical commutation relations to be
\bea
&&[A^a_i(\vec{x}),p^b_j(\vec{y})]=i\delta_{ij}\delta^{ab}\delta(\vec{x}-\vec{y});\nonumber\\
&&[p^a_i(\vec{x}),p^b_j(\vec{y})]=i\theta_{ij}\delta^{ab}\delta(\vec{x}-\vec{y});\nonumber\\
&&[A^a_i(\vec{x}),A^b_j(\vec{y})]=0.
\eea
Our aim is to deform the secondary constraint $\Delta^b$ in a manner preserving the gauge transformations (\ref{gatra}). It is easy to see that this can be achieved if we modify the secondary constraint as
\bea
\tilde{\Delta}^b=-(\pa_i p_i^b+gf^{bcd}A^c_i p_i^d)+\theta_{ij}(\pa_iA^b_j+\frac{1}{2}gf^{bcd}A^c_iA^d_j).
\eea
This modification of the secondary constraint implies in the modification of the Hamiltonian which acquires the form
\bea
\label{h}
\tilde{H}=\frac{1}{2}p^a_ip^a_i+\frac{1}{4}F^a_{ij}F^a_{ij}+A_0^b\theta_{ij}(\pa_iA^b_j+\frac{1}{2}gf^{bcd}A^c_iA^d_j).
\eea
Then, we can introduce the canonical momenta
\bea
\label{cm}
\pi_i^a=p_i^a-\frac{1}{2}\theta_{ij}A_j^a,
\eea
and they satisfy the commutation relation $[\pi_i^a,\pi_j^b]=0$.

The new Lagrangian is
\bea
\tilde{L}=\pi_i^a\dot{A}_i^a-\tilde{H}.
\eea
Substituting the canonical momenta (\ref{cm}) and the modfified Hamiltonian (\ref{h}) to this expression, we find that the new Lagrangian can be written as
\bea
\tilde{L}=L+\Delta L\equiv L-\frac{1}{2}\theta_{ij}\dot{A}_i^aA_j^a-A_0^b\theta_{ij}(\pa_iA^b_j+\frac{1}{2}gf^{bcd}A^c_iA^d_j).
\eea
As a result, we find
\bea
\Delta L=\theta_{ij}(-\frac{1}{2}\dot{A}_i^aA_j^a-A_0^a\pa_iA_j^a+\frac{1}{2}gf^{bcd}A_0^bA_i^cA^d_j).
\eea
After an appropriate symmetrization, introducing $\theta_{ij}=\epsilon_{0ij}\theta$, we find
\bea
\Delta L=\frac{1}{2}\theta\epsilon^{\mu\nu\lambda}(A_{\mu}^a\pa_{\nu}A_{\lambda}^a+\frac{1}{3}gf^{abc}A_{\mu}^aA_{\nu}^bA_{\lambda}^c)=
\frac{1}{2}\theta\epsilon^{\mu\nu\lambda}{\rm tr}(A_{\mu}\pa_{\nu}A_{\lambda}+\frac{2}{3}gA_{\mu}A_{\nu}A_{\lambda}),
\eea
which reproduces the structure of the well known non-Abelian Chern-Simons term, with the mass is proportional to the noncommutativity parameter, just as in \cite{NPR}.

We can try to implement a more general deformation of the canonical algebra, that is,
\bea
&&[A^a_i(\vec{x}),p^b_j(\vec{y})]=i\delta_{ij}\delta^{ab}\delta(\vec{x}-\vec{y});\nonumber\\
&&[p^a_i(\vec{x}),p^b_j(\vec{y})]=i\theta_{ij}\delta^{ab}\delta(\vec{x}-\vec{y});\nonumber\\
&&[A^a_i(\vec{x}),A^b_j(\vec{y})]=i\tilde{\theta}_{ij}\delta^{ab}\delta(\vec{x}-\vec{y}).
\eea
Let us impose again a requirement that the gauge transformations should have the form (\ref{gatra}). First of all, since $\theta_{ij}$ and $\tilde{\theta}_{ij}$ are constants, we suggest from the beginning that $\theta_{ij}=\theta\epsilon_{ij}$, $\tilde{\theta}_{ij}=\tilde{\theta}\epsilon_{ij}$.

To do it, let us suggest the following form of the modified secondary constraint which is the most general expression of no higher than second order in canonical variables:
\bea
\Phi^{(2)b}&=&-\pa_i p_i^b+k_1gf^{bcd}A^c_i p_i^d+k_2\epsilon_{ij}\pa_iA^b_j+k_3\epsilon_{ij}\pa_ip^b_j+k_4\epsilon_{ij}gf^{bcd}A^c_ip^d_j+\nonumber\\&+&
k_5gf^{bcd}\epsilon_{ij}p^c_ip^d_j+k_6gf^{bcd}\epsilon_{ij}A^c_iA^d_j.
\eea
Here the coefficients $k_1\ldots k_6$ depend on $\theta,\tilde{\theta}$. 

The corresponding variations of the fields look like
\bea
\label{var}
\delta A^a_n&=&\{A^a_n,\Phi^{(2)b}\}\xi^b=\pa_n\xi^a-k_1gf^{abc}\xi^b(\tilde{\theta}\epsilon_{ni}p^c_i-A^c_n)-k_2\tilde{\theta}\pa_n\xi^a+k_3
\epsilon_{ni}\pa_i\xi^a+\nonumber\\&+&k_4 gf^{abc}\xi^b(\tilde{\theta}p^c_n-\epsilon_{ni}A^c_i)-2k_5g\epsilon_{ni}f^{abc}\xi^bp^c_i+2k_6gf^{abc}\tilde{\theta}\xi^bA^c_n;
\nonumber\\
\delta p^a_n&=&\{p^a_n,\Phi^{(2)b}\}\xi^b=\theta\epsilon_{ni}\pa_i\xi^a+k_1gf^{abc}\xi^bp^c_n+k_1\theta g f^{abc}\epsilon_{ni}\xi^bA^c_i-k_2\epsilon_{ni}\pa_i\xi^a-k_3\theta\pa_n\xi^a+\nonumber\\&+&k_4\epsilon_{ni}gf^{abc}\xi^bp^c_i+
k_4\theta gf^{abc}A^c_n\xi^b+2k_5gf^{abc}\theta \xi^b p^c_n-2k_6gf^{abc}\epsilon_{ni}\xi^bA^c_i.
\eea 
We want these transformations to reproduce (\ref{gatra}). For the variation of $A^a_n$ this requirement yields $k_3=0,k_4=0$, so, we will not consider these terms in the equation for $\delta p^a_i$. Also, we find
\bea
k_2\tilde{\theta}=0;\quad\,k_1+2k_6\tilde{\theta}=-1,\quad\,k_1\tilde{\theta}+2k_5=0.
\eea 
For the second equation, after substituting $k_3=k_4=0$, we get
\bea
k_2=\theta, \quad\,k_1+2k_5\theta=-1,\quad\,k_1\theta-2k_6=0.
\eea
Comparing these equations, we find that the variations of the fields (\ref{var}) reproduce the  form of variations under the gauge transformations if and only if $\theta\tilde{\theta}=0$. Hence, we must have either $\tilde{\theta}=0$, which is exactly the case studied above, or $\theta=0$. Thus, we conclude that we cannot impose noncommutativity both in field and momentum sectors in a manner compatible with the gauge symmetry. 

It remains only to finish the study in the case when $\theta=0$. In this case, the modified constraint is
\bea
\Phi^{(2)b}&=&-\pa_i p_i^b-gf^{bcd}A^c_i p_i^d+
\frac{\tilde{\theta}}{2}gf^{bcd}\epsilon_{ij}p^c_ip^d_j,
\eea
and the modified Hamiltonian is
\bea
\label{h1}
\tilde{H}=\frac{1}{2}p^a_ip^a_i+\frac{1}{4}F^a_{ij}F^a_{ij}+A_0^b[-\pa_i p_i^b-gf^{bcd}A^c_i p_i^d+
\frac{\tilde{\theta}}{2}gf^{bcd}\epsilon_{ij}p^c_ip^d_j].
\eea
Since commutation relations between momenta are not modified in this case, the momenta $p^a_i$ continue to be canonical ones, whereas the coordinates -- do not more. The correct "new" canonical coordinates, whose commutators are equal to zero, are
\bea
\tilde{A}^a_i=A^a_i-\frac{1}{2}\tilde{\theta}\epsilon_{ij}p^a_j,
\eea
with the "old" velocities are related with momenta as
\bea
\dot{A}^b_i=\frac{\pa\tilde{H}}{\pa p^b_i}=p^b_i+\pa_iA^b_0+gf^{abc}A^a_0A^c_i+g\tilde{\theta}f^{abc}A^a_0\epsilon_{ij}p^c_j,
\eea
which for $\tilde{\theta}=0$ evidently reduces to the common expression (\ref{vel}).
Unfortunately, this equation, whose equivalent form is
\bea
p^c_j(\delta^{bc}\delta_{ij}+g\tilde{\theta}f^{abc}A^a_0\epsilon_{ij})=\dot{A}^b_i-\pa_iA^b_0+gf^{bac}A^a_0A^c_i\quad\,(=F^b_{0i}),
\eea
cannot be solved exactly, we can use only iterative approach (however, we would like to point out that this problem does not arises in the Abelian case where one finds $p^b_i=F^b_{0i}$). As a zeroth approximation (which, however, is sufficient to find the corrections in the effective Lagrangian up to the first order in $\tilde{\theta}$), we can use the $\tilde{\theta}=0$ expression for the canonical momentum $p^a_i$ (\ref{cm0}),
thus, the Lagrangian $\tilde{L}=p^a_i\dot{\tilde{A}}^a_i-\tilde{H}$ acquires a correction $\Delta L$ generated by modifications both of the Hamiltonian and $\dot{A}^a_i$. This correction, being expressed in terms of the canonical momenta, looks like:
\bea
\Delta L=-\frac{1}{2}\tilde{\theta}\epsilon_{ij}p^a_i\dot{p}^a_j-\frac{1}{2}gf^{bcd}A^b_0\tilde{\theta}\epsilon_{ij}p^c_ip^d_j.
\eea
This expresion is exact, without any approximations. After elimination of momenta, where we must employ the approximate expressions for $p^a_i$ in terms of velocities, we find that
\bea
\Delta L=-\frac{1}{2}\tilde{\theta}\epsilon_{ij}F^a_{0i}\dot{F}^a_{0j}-\frac{1}{2}gf^{bcd}A^b_0\tilde{\theta}\epsilon_{ij}F^c_{0i}F^d_{0j}+O(\tilde{\theta}^2).
\eea
Thus, one can see that, as a result, the modified Lagrangian in the case of noncommuting field operators involves higher derivatives (since $F^a_{0i}$ contain first temporal derivative). The similar conclusion, that is, generation of higher derivatives in the case of noncommuting fields (which can be treated as UV limit of the theory, see discussion of scales in the noncommutative fields method in \cite{Gamb0}), was obtained in \cite{Gamb}. Also, we note that, as this correction to the Lagrangian has quite ugly form, we can conclude that in this case, unlike of the case of noncommuting momenta, we meet an explicit Lorentz symmetry breaking.

Let us discuss the results. We studied the generalized version of the noncommutative field method, in which, differently from the most popular version \cite{Gamb1,Gamb2,ourgra} not only the commutation relations between canonical momenta are deformed but also the commutation relations between canonical field coordinates. The most important conclusions are the following ones. First, one cannot deform these two canonical commutation relations simultaneously in a manner compatible with the gauge symmetry. This fact can be treated as a need to choose between study of the low-energy behaviour (which corresponds to deformation of commutation relation between canonical momenta) and study of the high-energy behaviour (which corresponds to deformation of commutation relation between canonical fields) with no possibility to consider two limits at the same time. Second, in the low-energy limit the complete, non-linearized Chern-Simons term is generated, which is a natural non-Abelian generalization of the result obtained in \cite{NPR} where the quadratic Chern-Simons term was generated for the electrodynamics, with no Lorentz symmetry breaking terms arises in this case, and both the mass term and cubic interaction term with a correct coefficient are generated. However, the new term arisen in the high-energy limit turns out to break the Lorentz symmetry explicitly, and, moreover, it involves higher derivatives as it was predicted in \cite{Gamb}. The natural treating of this result is that the breaking of the Lorentz symmetry at high energies can be related to the GZK effect and many other studies predicting Lorentz symmetry breaking namely for high energy scales (see f.e. \cite{Mag}). 

{\bf Acknowledgments.}
The work by A. Yu. P. has been supported by CNPq-FAPESQ DCR program, CNPq project No. 350400/2005-9.


\begin{thebibliography}{50}

\bibitem{SW} N. Seiberg, E. Witten, JHEP {\bf 09}, 032 (1999), hep-th/9908142.

\bibitem{Gamb0} J. Carmona, J. Cortes, J. Gamboa and F. Mendez, 
Phys. Lett. B565, 222 (2003), hep-th/0207158;   JHEP {\bf 03}, 058 (2003), hep-th/0301248.

\bibitem{JK} R. Jackiw, V. A. Kostelecky, Phys. Rev. Lett. 82, 3572 (1999), hep-ph/9903158.

\bibitem{Gamb1} J. Gamboa, J. Lopez-Sarrion, Phys. Rev. D71, 067702 (2005), hep-th/0501034.

\bibitem{Gamb2} H. Falomir, J. Gamboa, J. Lopez-Sarrion, F. Mendez, A. J. da Silva, Phys. Lett. B632, 740 (2005), hep-th/0504032.

\bibitem{ourgra} A. F. Ferrari, M. Gomes, J. R. Nascimento, E. Passos, A. Yu. Petrov, A. J. da Silva, Phys. Lett. B652, 174 (2007),  hep-th/0609222.

\bibitem{NPR} J. R. Nascimento, A. Yu. Petrov, R. F. Ribeiro, Europhys. Lett. 77, 51001 (2007), hep-th/0601077.

\bibitem{Redlich} A. N. Redlich, Phys. Rev. D29, 2366 (1984).

\bibitem{DJT} S. Deser, R. Jackiw, S. Templeton, Ann. Phys. 140, 372 (1982); Phys. Rev. Lett. 48, 975 (1982).

\bibitem{quCS} A. P. Polychronakos, Phys. Lett. B241, 37 (1990); L.-S. Chen, G. Dunne, K. Haller, E. Lim-Lombridas, Phys. Lett. B348, 468 (1995).

\bibitem{Park} M.-I. Park, Nucl. Phys. B544, 377 (1999), hep-th/9811033.

\bibitem{Wo} L. S. Grigorio, M. S. Guimaraes, S. Wotzasek, "Induced deformation of the canonical structure and UV/IR duality in (1+1)D", arXiv: 0802.1193; J. Gamboa, L. S. Grigorio, M. S. Guimaraes, F. Mendes, S. Wotzasek, "Radiative processes as a condensation phenomenon and the physical meaning of deformed canonical structures", arXiv: 0805.0626.

\bibitem{Gamb} J. Gamboa, J. Lopez-Sarrion, A. Polychronakos, Phys. Lett. B634, 471 (2006), hep-ph/0510113.

\bibitem{Mag} J. Magueijo, L. Smolin, Class. Quant. Grav. 21, 1725 (2003), gr-qc/0305055.

\end{thebibliography}
\end{document}